%% file: napolitano.tex
\documentclass[graybox,natbib]{svmult}

\usepackage{mathptmx}       
\usepackage{helvet}         
\usepackage{courier}        
\usepackage{type1cm}        
\usepackage{makeidx}         
\usepackage{graphicx}        
\usepackage[bottom]{footmisc}
\usepackage[T1]{fontenc}

\usepackage{aas_macros}

\makeindex             

\def\atlas3d{ATLAS$^{\rm 3D}$}
\def\lsim{\mathrel{\rlap{\lower3.5pt\hbox{\hskip0.5pt$\sim$}}
    \raise0.5pt\hbox{$<$}}}                
\def\gsim{~\rlap{$>$}{\lower 1.0ex\hbox{$\sim$}}}

\begin{document}

\title*{Strong lens search  in the ESO public Survey KiDS}

\author{N.~R.~Napolitano, G.~Covone, N.~Roy, C.~Tortora, F.~La~Barbera, M.~Radovich, F.~Getman, M.~Capaccioli, A.~Colonna,  M.~Paolillo, G.~A.~Verdoes~Kleijn, L.V.E.~Koopmans and KiDS collaboration}
\authorrunning{Napolitano et al.}
\institute{ N.R.~Napolitano \email{napolita@na.astro.it},
C.~Tortora, F.~La~Barbera, F.~Getman \at INAF -- Osservatorio
Astronomico di Capodimonte, Salita Moiariello, 16, 80131 - Napoli,
Italy \and G.~Covone, N.~Roy, M.~Capaccioli, A.~Colonna,
M.~Paolillo \at Dipartimento di Scienze Fisiche, Università di
Napoli Federico II, Compl. Univ. Monte S. Angelo, 80126 - Napoli,
Italy \and M.~Radovich \at INAF -- Osservatorio Astronomico di
Padova-vicolo Osservatorio 5 - 35122 Padova,Italy \and
Gijs~A.~Verdoes~Kleijn, L.V.E.~Koopmans Kapteyn Astronomical
Institute, University of Groningen, P.O. Box 800, 9700 AV
Groningen, the Netherlands}

%
%
\maketitle

\abstract{We have started a systematic search of strong lens
candidates in the ESO public survey KiDS based on the visual
inspection of massive galaxies in the redshift range $0.1<z<0.5$.
As a pilot program we have inspected 100 sq. deg., which overlap
with SDSS and where there are known lenses to use as a control
sample. Taking advantage of the superb image quality of
VST/OmegaCAM, the colour information and accurate model subtracted
images, we have found 18 new lens candidates, for which
spectroscopic confirmation will be needed to confirm their lensing
nature and study the mass profile of the lensing galaxies.}


\section{Dark matter in galaxy cores} \label{sec:2}


The dominant role of cold dark matter (CDM) in shaping structures
in the Universe is now very well constrained and understood on the
largest scales, but less on the scales of individual galaxy
haloes. The assembly and evolution of CDM haloes can be studied in
detail with N-body simulations, which predict that the CDM density
profile, $\rho_{\rm CDM}(r)$, are well described by the so-called
NFW profile with $\rho_{\rm CDM}(r) \propto r^{-3}$ in the outer
regions, and $\rho_{\rm CDM}(r) \propto r^{-\alpha}$, in the
centre (with $\alpha =-1$, \cite{Navarro-1996}; $\alpha =-1.5$
\cite{Moore-1998}). However, N-body simulations only follow the
evolution of CDM particles, not including the complex physics of
gas and stars, which can alter the CDM profiles (e.g., via
adiabatic contraction, \cite{Gnedin-2004} ).

The study of the CDM profile in the inner regions of ETGs has
received much interest, due to the large amount of available data
samples (e.g., \cite{Hyde-2009}; \cite{Tortora-2012}). However, it
is made difficult by the degeneracy between the shape of the CDM
profile and that of the stellar Initial Mass Function (IMF;
\cite{Dutton-2012}; \cite{Tortora-2013}.)

\section{Strong Lensing to study the
mass density of galaxies}

Strong lensing (SL hereafter) is an ideal tool to perform accurate
studies of the CDM distribution of intermediate and massive
galaxies (e.g. with velocity dispersions, $\sigma_v> \rm 180~ km
\, s^{-1}$) independently of the dynamical status of the mass
within the arc features. SL studies, often in combination with
stellar dynamics in the galaxy central regions, have found that
the light and halo profiles conspire to have a total mass density
profile which is nearly isothermal (\cite{Koopmans-2006};
\cite{Bolton-2006}; \cite{Auger-2010}; \cite{Chae-2014};
\cite{Oguri-2014}), i.e. a total-mass density slope of $\alpha
\sim -2$. This ``conspiracy'', once thought to be universal, has
been recently found not to be so, as lower mass galaxies might
have steeper $\alpha$ (\cite{Dutton-2014}; \cite{Tortora-2014}).
The physical reasons for this systematic variation of the observed
slope resides in the way baryons and CDM are mixed, after having
been assembled over time \cite{Remus-2013}. Such detailed studies
have been performed over samples of nearby early-type galaxies and
a similar systematic study is ongoing at higher-$z$ SL systems
(e.g., by SLACS, \cite{Bolton-2008}; SL2S, \cite{Ruff-2011}),
where SL is much more effective than standard dynamical studies.
For this reason it is becoming of fundamental importance to
enlarge the sample of SL systems either in the galaxy parameter
space (e.g. toward less massive systems, like sub-$L_*$) and in
redshift space.

\section{Why galaxy-galaxy SL in the ESO Public Survey KiDS?}

The ESO public survey KiDS (Kilo Degree Survey, \cite{de
Jong-2013}) is a multi-band ($ugri$) imaging survey, designed to
take advantage of the superb seeing of Paranal and high image
quality of the VST/OmegaCAM. Its primary usage is weak lensing
measurements over 1500 sq. deg of the sky.

As such, KiDS is particularly suitable for  a systematic census of
lensing galaxies based on the identification of arc-like structure
from intermediate to massive galaxies (i.e., with velocity
dispersion of the order of  200 $\rm km \, s^{-1} $ or larger), to
galaxy groups and galaxy clusters, thus covering a wide range of
lens masses. Indeed, the Einstein ring writes easily as a function
of the velocity dispersion as $ R_{\rm E} \sim 1'' \times
(\sigma_v / 220 \, \rm km \, s^{-1})^{2}$, and for a typical $FWHM
\sim0.7''$ in $r$-band, we can expect to observe lensing arcs of
gravitational structures of $\sigma_v> \rm 180~ km \, s^{-1}$.

\begin{figure}
\centering
\includegraphics[scale=.35]{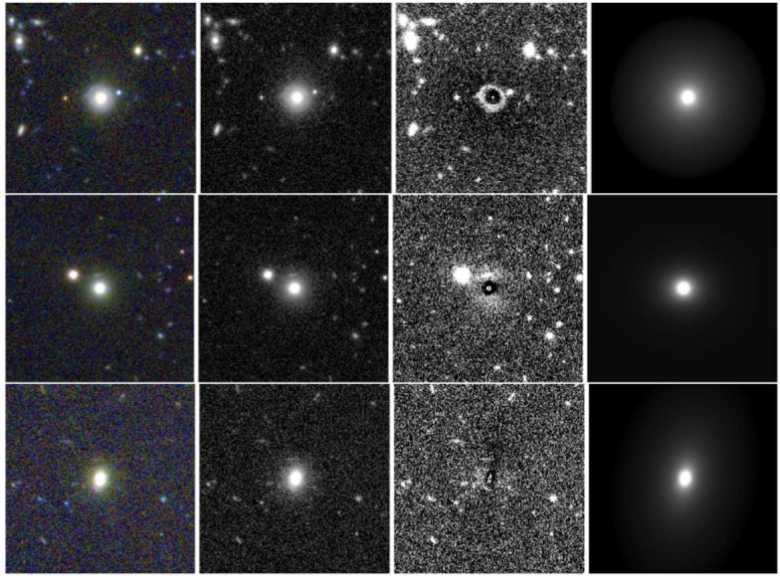}
\caption{Results from the visual inspection of SL candidates. Top
rows show two high-confidence candidates,  bottom is a a less
secure one. At each row we show: 1) RGB colour image, 2) $r-$band
image; 3) model subtracted image; 4) 2DPHOT galaxy model. In the
top row, an Einstein ring is evident and shows up clearly in the
residual image. In the middle row a blue arc is also evident in
the colour image and nicely spotted into the residual image. In
the bottom row a faint blue arc is seen in the colour image in the
bottom edge of the redder central galaxy, which shows up slightly
clearer in the model subtracted image. Estimated $R_E$ are $6-8$
arcsec.} \label{fig:1}
\end{figure}

The search for SL in KiDS is expected to provide up to thousands
galaxy rings (e.g. by extrapolating CFHT-LS numbers,
\cite{Oguri-2014}). This can be performed only with
(semi-)automated techniques (\cite{Alard-2007};
\cite{Gavazzi-2014}), as potential candidates in KiDS can be of
the order of thousands per square degree, making visual inspection
inapplicable. However, lensing finding techniques still suffer
from a number of issues yet to be solved, affecting the
completeness and reliability of their catalogs and need to be
trained over visually-inspected samples.

\section{Selection of SL candidates via visual inspection}

Due to the yet limited extension of the area (150 sq. deg.)
included in the KiDS ESO-Data Release 2 (ESO-DR2), we have decided
to start a visual search based on simple criteria: 1) sources with
SDSS spectroscopic redshifts between $z=0.1$ and $z=0.5$ and 2)
sources brighter than $r=20$, which together maximise the SL
probability for the depth allowed by KiDS images. These criteria
were restrictive and biased toward larger R$_E$ easier to spot by
visual inspection.
In order to quantify the performance of our search, we decided to
start on some control fields, overlapping with the SDSS, and
selected $\sim$600 lens candidates in $\sim$100 sq.deg, including
a ten of SLACS lenses.

The visual inspection (performed by six observers) was based on a)
the galaxy $r-$band image ($r-$band is chosen as reference since
it is the deepest and highest quality among KiDS wavebands), b) a
three-colour combined image to find a gradient in colour between
the candidate lens galaxy and the arc, c) the 2DPHOT model
\cite{LaBarbera-2008} subtracted image to detect residual arc
features. Examples of lens candidates are in Figure~\ref{fig:1}).
Following this procedure, we ended up with a list of 18 new
potential lens candidates, half of which have high significance. 

The spectroscopic follow-up will provide us with the confirmation
of the lensing nature of the lens candidates and the independent
kinematic information to constrain the mass profile of the lensing
galaxies.

\input{lensing_referenc}

\end{document}

%% file: lensing_referenc.tex
\bibliographystyle{plain}   

%% file: napolitano.bbl
\begin{thebibliography}{50}

\bibitem{Alard-2007} Alard, C.\ 2007, \mnras, 382, L58

\bibitem{Auger-2010} Auger, M.~W., Treu, T., Bolton, A.~S., et al.,\ 2010, \apj, 724, 511

\bibitem{Bolton-2006} Bolton, A.~S., Burles, S., Koopmans, L.~V.~E., Treu, T., Moustakas, L.~A.,\ 2006, \apj, 638, 703

\bibitem{Bolton-2008} Bolton, A.~S., Burles, S., Koopmans, L.~V.~E., et al.\ 2008, \apj, 682, 964

\bibitem{Chae-2014} Chae, K.~H., Bernardi, M., Kravtsov, A.~V.,\ 2014, \mnras, 437, 3670

\bibitem{de Jong-2013} de Jong, J.~T.~A., Kuijken, K., Applegate, D., et al.,\ 2013, The Messenger, 154, 44

\bibitem{Dutton-2012} Dutton, A.~A., Mendel, J.~T., Simard, L.,\ 2012, \mnras, 422, L33

\bibitem{Dutton-2014} Dutton, A.~A., \& Treu, T.,\ 2014, \mnras, 438, 3594

\bibitem{Gavazzi-2014} Gavazzi, R., Marshall, P.~J., Treu, T., Sonnenfeld, A.,\ 2014, \apj, 785, 144

\bibitem{Gnedin-2004} Gnedin, O.~Y., Kravtsov, A.~V., Klypin, A.~A., Nagai, D.,\ 2004, \apj, 616, 16

\bibitem{Hyde-2009} Hyde, J.~B., \& Bernardi, M.,\ 2009, \mnras, 396, 1171

\bibitem{Koopmans-2006} Koopmans, L.~V.~E., Treu, T., Bolton, A.~S., Burles, S., Moustakas, L.~A.,\ 2006, \apj, 649, 599

\bibitem{LaBarbera-2008} La Barbera, F., de Carvalho, R.~R., Kohl-Moreira, J.~L. et al.,\ 2008, \pasp, 120, 681

\bibitem{Moore-1998} Moore, B., Governato, F., Quinn, T., Stadel, J., Lake, G.,\ 1998, \apjl, 499, L5

\bibitem{Navarro-1996} Navarro, J.~F,  Frenk, C.~S., White, S.~D.~M.,\ 1996, \apj, 462, 563

\bibitem{Oguri-2014} Oguri, M., Rusu, C.~E., Falco, E.~E.,\ 2014, \mnras, 439, 2494

\bibitem{Remus-2013} Remus, R.~S., Burkert, A., Dolag, K., et al.\ 2013, \apj, 766, 71

\bibitem{Ruff-2011} Ruff, A.~J., Gavazzi, R., Marshall, P.~J., Treu, T., Auger,M.~W., Brault, F.,\ 2011, \apj, 727, 96

\bibitem{Tortora-2012} Tortora, C., La Barbera, F., Napolitano, N.~R., de Carvalho, R.~R., Romanowsky, A.~J.,\ 2012, \mnras, 425, 577

\bibitem{Tortora-2013} Tortora, C.,  Romanowsky, A.~J., Napolitano, N.~R.,\ 2013, \apj, 765, 8

\bibitem{Tortora-2014} Tortora, C., La Barbera, F., Napolitano, N.~R., Romanowsky, A.~J., Ferreras, I., de Carvalho, R.~R.,\ 2014, \mnras, 445, 115


\end{thebibliography}
